# Ferromagnetism in 2D Vanadium Diselenide

*Xiong Wang [1], Dian Li [1], Zejun Li [2], Changzheng Wu [2], Gang Chen [1], Xiaodong Cui [1]\**

1. Physics Department, University of Hong Kong, Hong Kong, China

2. Hefei National Laboratory for Physical Sciences at the Microscale, CAS Center for Excellence in Nanoscience, and CAS Key Laboratory of Mechanical Behavior and Design of Materials, University of Science and Technology of China, Hefei, China

\*e-mail: xdcui@hku.hk

**Two-dimensional (2D) Van der Waals ferromagnets carry the promise of ultimately miniature spintronics and information storage devices.[1,2] Among the newly discovered 2D ferromagnets all inherit the magnetic ordering from their bulk ancestors.[3-5] Here we report a new 2D ferromagnetic semiconductor at room temperature, *2H* phase vanadium diselenide ($VSe_2$) which show ferromagnetic at 2D form only. This unique 2D ferromagnetic semiconductor manifests an enhanced magnetic ordering owing to structural anisotropy at 2D limit.**

Magnetism has been a long-lasting fascinating topic in condensed matter physics and been carrying a potential for next generation informatics and electronics[1,2]. Although the Mermin-Wagner theorem indicates the absence of the long-range magnetic order for the two-dimensional (2D) isotropic Heisenberg model at the finite temperatures[6], magnetic order still could survive in the anisotropic 2D systems where symmetry breaking is materialized either by the finite system size, defect effects (breaking the lattice translation symmetry) or the anisotropic spin exchange interactions (breaking the spin rotational symmetry). With the growing enthusiasm in the emerging Van der Waals crystals, 2D ferromagnetic materials including $Cr_2Ge_2Te_6$, $CrI_3$ and $Fe_3GeTe_2$ have been discovered and construct a new family of 2D Van der Waals ferromagnets[3-5]. As yet these newly discovered 2D magnets inherit the magnetic properties of their bulk crystals as the magnetic order sustains from multiple layers down to monolayers. Although monolayer *1T*-$VSe_2$ with ferromagnetism was theoretically predicted and experimentally reported recently[7-11], the controversial results with both theoretical and experimental evidences raised ambiguity[12-16]. Here we study the magnetic properties of the 2D $VSe_2$ single crystal samples with the polar magnetic circular dichroism (MCD) microscopy and the second harmonic generation (SHG) technique. A strong ferromagnetism is observed in the *2H* phase of $VSe_2$ with a Curie-Weiss temperature up to 425 K, and this ferromagnetism softens with the increased sample thickness. We attribute the change of the magnetic properties to the changing of the magnetic anisotropy that fundamentally arises from the spin-orbit coupling. Our finding provides a new ferromagnetic semiconductor with a potential of building blocks for spintronics and valleytronics and a microscopic understanding of the internal interaction in 2D ferromagnets.

Vanadium diselenide is among the family of transition metal dichalcogenides (TMDCs) which are believed to be paramagnetic and feature strong light-matter interactions, spin and valley degrees of freedom and giant spin-valley coupling at the 2D limit. Most TMDC materials show the polymorphism and could exist in two different structural phases as illustrated in Fig 1a, the *2H* phase with a trigonal prismatic cell and a point group symmetry $D_{6h}^4$ and the *1T* phase with an octahedral cell and a point group symmetry $D_{3d}^3$, respectively [17]. In contrast to its celebrated sister compounds $MoS_2$ and $WSe_2$, the $VSe_2$ crystals usually exist in the *1T* form instead of *2H* [18]. While the dimensionality decreases, the 2D $VSe_2$ would thermodynamically favors the *2H* phase and the structural phase transition from *1T* to *2H* phase could be irreversibly materialized by annealing[19]. Vanadium is among the group *V* element with electron configuration of *$3d^3 4s^2$* and is well-known for the strong electron correlation owing to its 3*d*-shell orbits. In the unit cell of monolayer *H* phase $VSe_2$, the Se-V-Se bonds yield one *d*-shell electron per cell for the $V^{4+}$ ion, making $VSe_2$ a promising candidate of the Mott insulator with ferromagnetism. The spin-orbit coupling is usually active for the *$d^1$* electron configuration of the $V^{4+}$ ions, and our experimental study here does indicate the importance of the spin-orbit coupling in the understanding of the magnetic properties.

The phase transition occurs in $VSe_2$ from *1T* to *2H* by annealing the multi-layer flake, accompanying with the metal-insulator transition as shown in Fig. 1b & 1c.[19] Fig. 1d summarizes the MCD data of the *1T* phase and *2H* phase 2D $VSe_2$ at the room temperature. The *1T* phase $VSe_2$ both at bulk form and 2D cases (down to 20-layer) shows a paramagnetic response which is consistent with the recent report[12] where no ferromagnetism was observed at *1T* $VSe_2$ monolayer down to 10 K. In contrast, robust ferromagnetism displays in the *2H* phase $VSe_2$ multilayers, which is consistent with the previous calculations[20-22]. These contrasting magnetic behaviors fit the intuition of solid-state physics. Although the $V^{4+}$ ion carries single electron, for such a $VSe_2$ compound, ferromagnetism most likely is realized via super-exchange interaction and magnetic anisotropy. At each unit layer, the inversion symmetry is explicitly broken in *2H* phase. This inversion symmetry breaking induces a spin-orbit coupling as well as a charge band gap. The mirror symmetry with respect to the vanadium atom plane and the three-fold of rotation symmetry secure the spin orientation along the out-of-plane direction. In addition, the partially filled *d*-orbital electron shell allows the atomic spin-orbit coupling to be active. These factors increase the magnetic anisotropy of the *2H* phase against the *1T* phase and explain the magnetic difference between *2H* and *1T* phases.

Ferromagnetism in the *2H*-$VSe_2$ clearly shows a thickness dependence. It monotonically softens with the increased sample thickness within the sample range where the reproducible *2H* phase samples span from 15 nm (about 20 layers) to 47 nm in our experiments, until the ferromagnetism disappears at the sample thicker than 50 nm. The thickness dependent coercive force and the saturated MCD signal are summarized in Fig. 2b & 2c.

The thickness dependence could also be attributed to the magnetic anisotropy. There could be two kinds of anisotropy playing a role in the ferromagnetism here. The first lies in the layer-layer stacking order. The *2H* phase crystals follow a Bernal packing order and the *A-B-A* stacking restores the spatial inversion symmetry as a whole in the Van der Waals crystals. As the thickness thins into 2D, the inversion symmetry is gradually broken at multilayer level. This is demonstrated by the characterization of second harmonic generation (SHG) which is a second order nonlinear optical effect displayed in systems without structural inversion symmetry. Fig. 3b presents a SHG mapping on *2H* VSe$_2$ nano-flakes. The SHG intensity monotonically increases with the reduced thickness, implying the increased spatial asymmetry at thinner crystals. The other origin of the anisotropy is the interface effect. The interface breaks the inversion symmetry that again induces magnetic anisotropy and renders the out-of-plane easy axis for ferromagnetic thin films. For the 2D VSe$_2$, the anisotropy owing to interface effect dramatically increase as the sample shrinks to atomic layers.

The saturated MCD signals well follow the power-law form of $\alpha(1 - T/T_C)^\beta$ where $\alpha$, $\beta$ and $T_C$ as simultaneous fitting parameters. For the representative multilayer sample of a 16.9 nm thick *2H*-VSe$_2$ flakes, $\beta$ is given at 0.332 ± 0.045, consistent with $\beta = 0.326$ for the 3D Ising model (Adj. R-Square is 0.995). The deviation from the Heisenberg universality class indicates the presence of the magnetic anisotropy that lowers the spin symmetry from *SO(3)* down to *Z2*. The Ising universality class is consistent with the expectation for the spin-orbit coupling that breaks the spin rotational symmetry and generates the magnetic anisotropy.

In summary, we demonstrated the *2H* phase VSe$_2$ multilayers as a high temperature 2D ferromagnetic semiconductor. The ferromagnetic ordering exists only at 2D form owing to the enhanced structural anisotropy.

**Methods**

**Sample preparation.** The chemical vapor transport method was used to grow VSe$_2$ single crystals with iodine as the transport agent. The typical method is to put a mixture of vanadium, selenium powder and iodine powder with a stoichiometric ratio of 1:2 into a vacuum quartz tube. Then, the tube was put into a muffle furnace. Warm up to 850 °C, store for 2 days, slowly cool to 500 °C within 2 days, and finally cool to room temperature. A few millimeters of VSe$_2$ crystals can be obtained.

The VSe$_2$ single crystals are grown using chemical vapor transport method and the nanoflakes with the thickness ranging from about 10 to 50 nm are mechanically exfoliated onto silicon substrates capped with a 300-nm oxide layer in a glove box (H$_2$O and O$_2$ < 0.1 ppm). We cannot obtain samples thinner than 8 nm in our mechanical exfoliation. The thickness of flakes was visually pre-screened under optical microscope and precisely measured with atomic force microscope (AFM). The thermal annealing was carried out in a tube furnace in argon atmosphere. After annealing at 650 K, the samples were cooled down to ambient temperature in the same argon environment. The sample was placed in a high vacuum chamber (about $10^{-6}$ mbar) to prevent the influence of air and moisture.

**MCD microscopy.** Materials with non-zero magnetic moments can exhibit magnetic circular dichroism (MCD). MCD arises from the differential absorption of left-circularly polarized (LCP) light and right-circularly polarized (RCP) light. The MCD measurements were performed at a home-made heating stage in a vacuum with a temperature range from 300 K to 425 K under an out-of-plane magnetic field. An incident beam from the power-stabilized 633 nm laser diode of about 80 µW was parallel to the magnetic vector and normal to the reflecting surface. The beam was focused through an objective lens onto the $VSe_2$ flakes with a spot size of about 1 µm and the reflected beam was collected by the same objective lens. The incident beam was modulated by a photoelastic modulator (PEM). Then the MCD signal carried by the reflected beam was detected by balanced photodiode.

**SHG measurement.** The second harmonic generation (SHG) measurement was carried out by an excitation pulse from a Ti: sapphire oscillator (120fs, 80MHz).

**Acknowledgement:**

The work was supported by Croucher foundation, GRF #17304518, CRF C7036-17W of the Research Grant Council of Hong Kong and MOST 2020YFA0309603.

**Author Contribution**

XC conceived and supervised the project. XW and DL conducted the experiments and analyzed the data. ZL and CW fabricated the $VSe_2$ single crystals. GC provided the theory support. XW, GC and XC wrote the paper.

**Additional information**

The authors declare no competing financial interests. Correspondence and requests for materials should be addressed to X.C.

**Figures and figure captions**

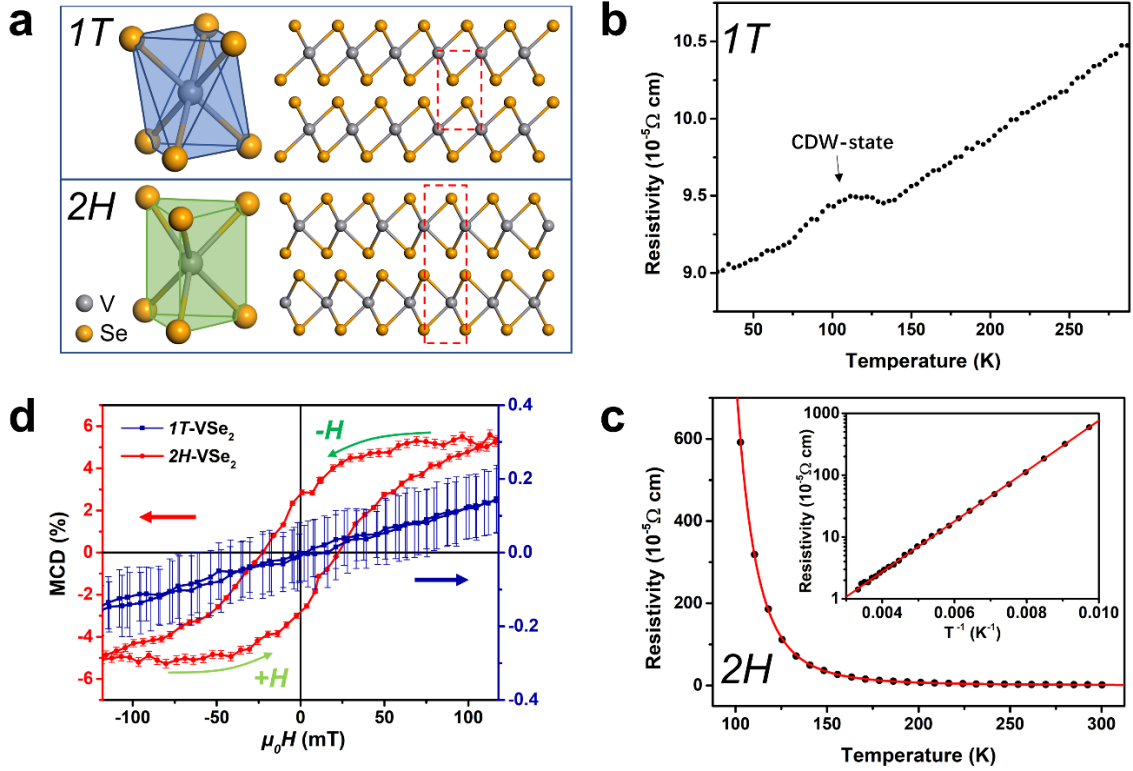

**Figure 1. Comparisons of structural symmetry, electrical and magnetic properties between two phases of 2D VSe$_2$. a,** The crystal structures of the *1T*- and *2H*-VSe$_2$. **b, c,** The crystal unit cells are represented by the red dashed rectangular boxes. The representative temperature dependent electric resistivity of *1T*-VSe$_2$ (**b**) and *2H*-VSe$_2$ (**c**). The temperature dependence shows the contrasting behaviors: metallic *1T* phase vs. semiconducting *2H* phase correspondingly. **d,** Representative MCD signals for the corresponding multilayer VSe$_2$ at 300 K. Error bars indicate the standard deviation (SD) of sample signals.

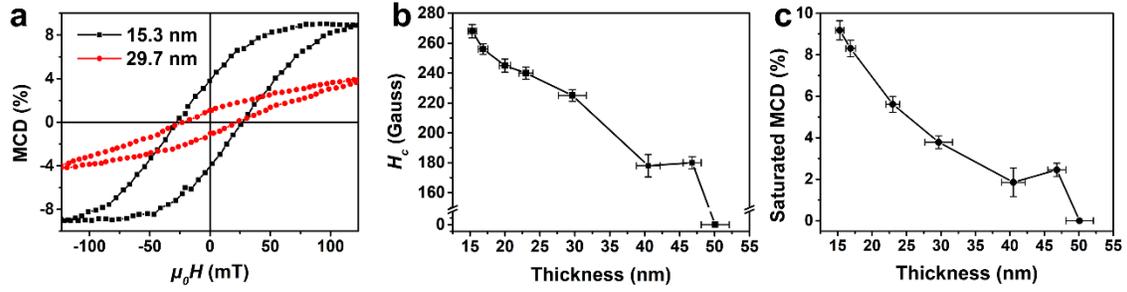

**Figure 2. Magnetic properties of multilayer *2H*-VSe$_2$ on Si/SiO$_2$ substrates at different thicknesses**. **a**, MCD hysteresis loops of *2H*-VSe$_2$ of 15.3 and 29.7 nm, respectively. **b, c,** The thickness dependence of coercive field $H_c$ (**b**) and saturated MCD signal (**c**) for *2H*-VSe$_2$. The $H_c$, MCD signal and thickness error bars indicate uncertainties in calibration of measurements.

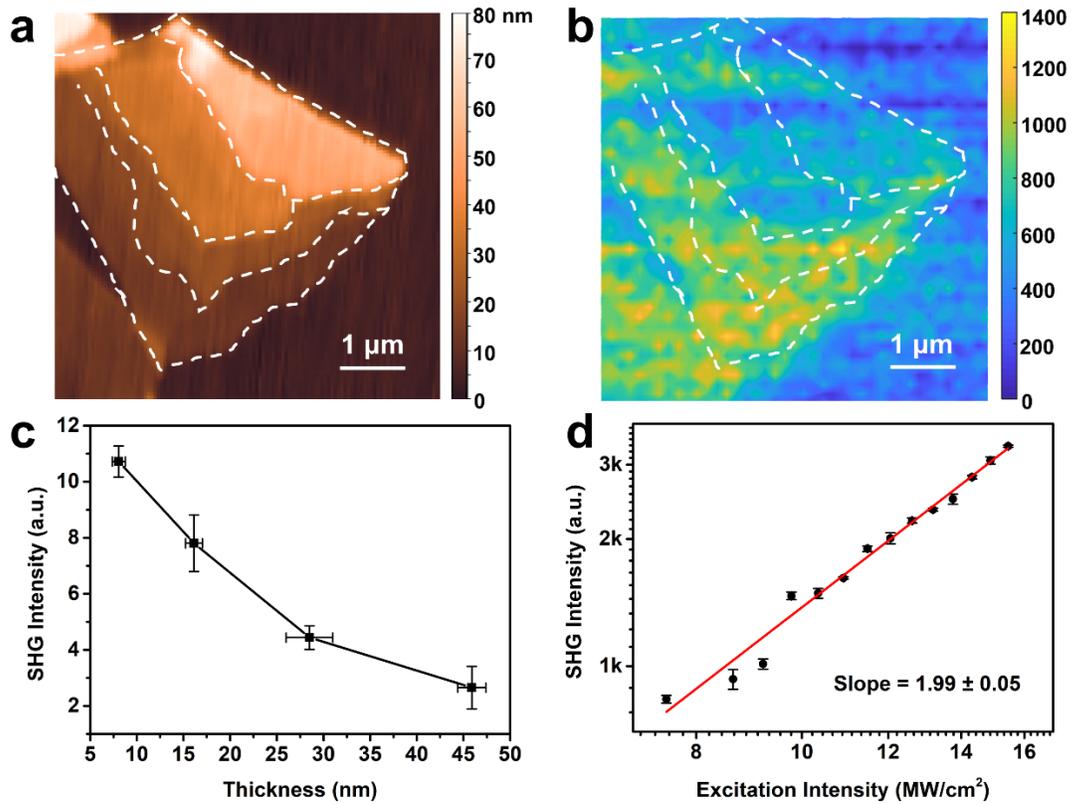

**Figure 3. The change of inversion symmetry with various thickness of *2H*-VSe$_2$. a, b,** The AFM topography (**a**) and the SHG mapping (**b**) of the *2H*-VSe$_2$ nano-flakes. **c,** The sample thickness dependence of SHG intensity for multilayer *2H*-VSe$_2$. The error bars correspond to the standard deviations of SHG signals from four regions of various thickness. **d**, The SHG intensity as a function of the excitation intensity.

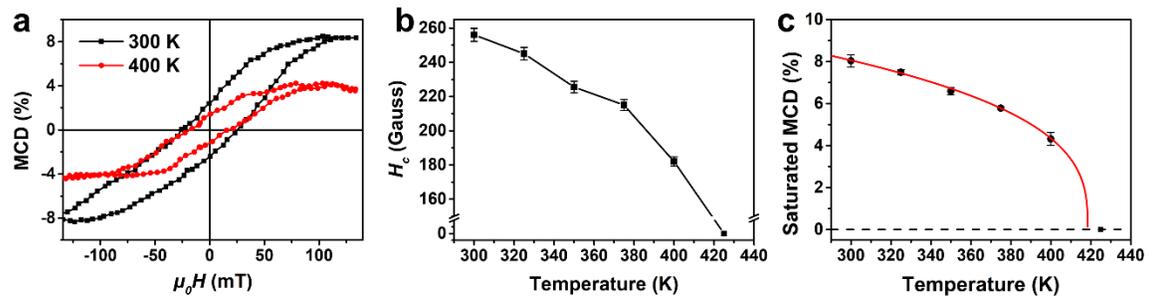

**Figure 4. The representative temperature dependent MCD of a 16.9 nm-thick *2H*-VSe₂. a,** The hysteresis loops of the *2H*-VSe₂ at 300 and 400 K. **b, c,** The coercive field $H_c$ (**b**) and the saturated MCD signal (**c**) as a function of temperature. The fitting curve follows the form of $\alpha(1 - T/T_C)^\beta$ where $T_C$ = 418.5 ± 7.8 K is extracted.